\documentclass[letterpaper]{article}
\usepackage{physics}
\usepackage{amsmath}
\usepackage{natbib,alifeconf}  

\title{Can justice be fair when it is blind? How social network structures can promote or prevent the evolution of despotism}

\author{Cedric Perret$^{1}$, Simon T. Powers$^1$, Jeremy Pitt$^{2}$  \and Emma Hart$^{1}$ \\
\mbox{}\\
$^1$Edinburgh Napier University, Edinburgh, EH10 5DT \\
$^2$Imperial College London, London, SW7 2BT \\
c.perret@napier.ac.uk, s.powers@napier.ac.uk, e.hart@napier.ac.uk, j.pitt@imperial.ac.uk} 

\begin{document}
\maketitle

\begin{abstract}
Hierarchy is an efficient way for a group to organize, but often goes along with inequality that benefits leaders. To control despotic behaviour, followers can assess leaders' decisions by aggregating their own and their neighbours' experience, and in response challenge despotic leaders. But in hierarchical social networks, this interactional justice can be limited by (i) the high influence of a small clique who are treated better, and (ii) the low connectedness of followers. Here we study how the connectedness of a social network affects the co-evolution of despotism in leaders and tolerance to despotism in followers. We simulate the evolution of a population of agents, where the influence of an agent is its number of social links. Whether a leader remains in power is controlled by the overall satisfaction of group members, as determined by their joint assessment of the leader’s behaviour. We demonstrate that centralization of a social network around a highly influential clique greatly increases the level of despotism. This is because the clique is more satisfied, and their higher influence spreads their positive opinion of the leader throughout the network. Finally, our results suggest that increasing the connectedness of followers limits despotism while maintaining hierarchy.

\end{abstract}

\section{Introduction}
The efficiency and stability of self-organized open systems are strongly dependent on the agents' capacity to sustain coordination and cooperation. For instance, to build a new structure, a group of robots need to decide of their respective roles such as harvester to get resources, manufacturer to create the required parts and builder to assemble them. Power suppliers and consumers within a smart grid have to adjust the scheduling of production and consumption to efficiently reduce power waste. Humans collectively decide of rules to manage common resources and limit selfish behaviours. Whether it is in artificial or natural social systems, autonomous agents need to constantly take collective decisions to coordinate complex tasks. Furthermore, collective decisions can also encompass the creation and modification of crucial institutional rules that affect group welfare e.g. rules governing the distribution of resources; as in human societies \citep{Ostrom1990GoverningCommons,North1990InstitutionsPerformance} or electronic self organized institutions \citep{Pitt2012AxiomatizationInstitutions}.  
To cope with the complexity inherent of large scale coordination, human societies tend to facilitate collective decision making by switching to hierarchy with a minority of influential individuals i.e. leaders, and a majority of influenceable individuals i.e. followers \citep{Hooper2010AGroups,Powers2014AnDespotism,Perret2017EmergenceModel}. Similarly, hierarchy has been shown to be highly relevant to artificial social systems by reducing the cost of organization \citep{Pugliese2015EmergenceRobots,Chih-Han2010CollectiveLeadership}.
However, hierarchy also implies an important cost. In human societies, leaders tend to evolve despotic behaviour where they exploit followers in order to increase their own resources \citep{Mitchell1915PoliticalDemocracy,Carneiro1970AState,Hayden2010WhoFutuna}. This ``iron law of oligarchy'' ultimately leads to an overall reduction in the productivity of the group (e.g less production of resources through collective action) \citep{Summers2005TheDespotism}. Evidently, such behaviour should be avoided in an artificial social system. Yet, the mechanisms behind the evolution of such despotism and inequality are still not fully understood.

Common explanations emphasize the importance of a surplus of transmissible resources \citep{Mattison2016TheInequality}, the capacity of followers to avoid domination, or the ability of leaders to impose domination \citep{Johnstone2000ModelsSynthesis,Summers2005TheDespotism}. However, and as described by the famous quote of John Acton: \textit{``Power tends to corrupt, and absolute power corrupts absolutely''}, it has also been suggested that the sole asymmetrical distribution of power is enough to lead to inequality and despotism.  \citep{Mitchell1915PoliticalDemocracy}. Power can be defined as the influence of individuals on collective decisions. Leaders might exert this influence to leverage institutional rules, and ultimately tilt the distribution of cost and benefits toward their own advantage.
But even if leaders have a huge influence on collective decisions in hierarchical societies \citep{Gavrilets2016ConvergenceLeadership}, their behaviours are constrained indirectly by the satisfaction of the rest of the group. For instance, an intolerable selfish leader could be, for the followers, worth the cost of overthrowing him. However, followers often lack direct knowledge of the leader's behaviours and decisions. To enforce such control,  they judge decision-makers by the state of the laws and rules they manage. In response to too despotic rules, they can then start a revolution to overthrow the leader.
This form of justice, where individuals judge how institutions and decision-makers treat them, is defined as interactional justice \citep{Schermerhorn2012OrganizationalBehavior}, and is a common way in which individuals exert control over their institutions in natural social systems. 
Yet, the monitoring of institutions and \textit{a fortiori} leaders is greatly dependent on the individuals having knowledge of the state of the system. This is because the agents first need to make a self-assessment of how they are being treated, i.e. build their own opinion from their personal experiences; and then make a collective assessment about whether to try to induce change, i.e. aggregate the opinions of other agents. Although this kind of opinion formation based on individual and social learning produces a global assessment of the current state of institutions, it is also constrained by the knowledge accessible to individuals. Because of the size and complexity of large human groups, this knowledge is often incomplete. Indeed, leaders have a strong influence on followers' opinion. In addition, leaders are often surrounded by a clique -- a limited number of highly influential individuals, such as a patriarchal clan in early agricultural societies \citep{Kaplan2009TheOrganization} or key policy-makers in contemporaneous communities \citep{Miller1958Decision-makingCity}. By providing them with preferential access to resources, leaders can cause the clique to have a positive opinion, which they then spread throughout the network as a result of their high connectedness. The opinion of followers thus becomes biased by the clique, blinding them to the actual level of inequality. This blindness limits the control of followers on the leader's decision.
In recent work, \citet{Pitt2017InteractionalSystems} modelled this process and has formally demonstrated that a centralized social network with a leader and a clique biases the transfer of knowledge, and ultimately leads to misconceptions on the current fairness level of the society. From this, it was predicted that an incomplete transfer of knowledge could blind the interactional justice of followers and allow the evolution of despotic leaders. However, this prediction has remained untested so far.

Here we address this by using an evolutionary model to answer the following question:
\textit{how does social network structures affect the evolution of despotism in hierarchical societies ?} In the model, agents are explicitly organized in a centralized network with the leader as the central node. Agents are described by their preferences on the distribution of resources, and their opinions on the actual level of fairness in the society. We use a Moran process \citep{Moran1958RandomGenetics,Lieberman2005EvolutionaryGraphs} to simulate the evolution of their distribution preferences and study how the network structure affects the level of despotism. Our results highlight a negative effect of centralization on the evolution of distribution preferences, with more centralised networks leading to followers accepting higher levels of despotism as interactional justice becomes biased. But on the flip side, we show that the level of despotism can be limited by increasing the connectedness of followers. By doing so, interactional justice becomes effective at constraining the leader, allowing the benefits of hierarchy to be realized without the costs. Our results contribute to the knowledge on the evolution of hierarchy, institutions and justice, key concepts to understand natural social systems and design artificial self-organized system. 

\section{Model definition}
To investigate the impact of social network structure on the evolution of despotism, we have developed a model to simulate the evolution of distribution preferences within a hierarchical society. This section provides an outline of the model with a detailed description of the mechanisms implemented: the network structure, the distribution of resources, the interactional justice and the reproduction.  

\subsection{Model outline and life cycle}
We consider a fixed-size population of $N$ individuals explicitly organized in a directed network. The population is composed of one leader deciding of the distribution of resources and $N- 1$ followers. In addition, the population is divided between $N_c$ highly influential individuals called clique members which includes the leader, and $N_o$ individuals with low influence called outgroup members . 
The life cycle consists of: 
\begin{enumerate}
\item The group produces an amount of resource that is distributed amongst group members according to the distribution preference of the leader, $z_L$ (Equation~\ref{eqnDistrib}).
\item Each individual builds its \emph{own subjective mindset} about the fairness of resource allocation, $m$, as a function of the resources it personally received and its own distribution preference (Equation~\ref{eqnMindset}).
\item Each individual builds its \emph{opinion} about the overall fairness of the resource distribution, $o$, by aggregating its own mindset and the mindsets of the neighbouring individuals (individuals linked to the focal individual) (Equation~\ref{eqnOpinion}).
\item Each individual compares its \emph{opinion} to its distribution preference $z$. If the opinion is higher than the preference, the individual is considered defiant and pays a cost to attempt a revolution.
\item In case of a large proportion of defiant individuals within the population, i.e. above a revolution threshold $T$, a new leader and clique are chosen within the defiant individuals. The network is then rebuilt.
\item A random individual dies and is replaced by another individual with a probability proportional to its fitness (Equation~\ref{eqnRepro}). This reproductive process is repeated $R$ times.
\end{enumerate}

Individuals are modelled by one cultural trait; their distribution preference $z$ defined between equal ($z=0$) and strongly skewed ($z=1$). In leaders, this trait $z_L$ is translated into the function defining the distribution of resources with $z_L=0$ representing a fair leader and $z_L=1$ a despotic leader. In followers, this trait is translated into their tolerance towards inequality, with the minimum $z=0$ equal to no tolerance and the maximum $z=1$ equal to the maximum tolerance where any level of despotism is accepted. 
The agents are also indirectly described by their influence $\alpha$, here translated into the probability that the focal node is connected toward another individual. The trait $z$ evolves following a Moran process described in the reproduction section \citep{Moran1958RandomGenetics,Lieberman2005EvolutionaryGraphs}. 
In addition, when a new individual is born, its $z$ trait can mutate at a rate $\mu_\mathrm{z}$. When a mutation occurs, a random value is sampled from a truncated Gaussian distribution centered on the current value of the trait, with variance $\sigma_\mathrm{z}$.

\subsection{Network structures}
To study how hierarchy can affect the evolution of despotism, we explicitly describe the social structure of the population by a directed social network. In this network, each node represents an individual and each directed link represents a social contact from one individual to another. We define the in-degree and the out-degree of an individual as the number of links connected respectively toward and from this focal node. The influence of an individual is taken to be its connectedness $\alpha$ defined as the probability of an individual to be connected toward another individual.
To build the network, we use an algorithm derived from the Erd\H{o}s--R\'{e}nyi model \citep{Erdos1959OnI} as follows: 
\begin{enumerate}
\item The leader and all members of the clique are fully connected.
\item For each individual, a directed link is created from the individual $i$ to the individual $j$ following a probability $\alpha_c$ if the focal individual $i$ is member of the clique or is the leader, and $\alpha_o$ if the focal individual $i$ is a follower.
\end{enumerate}
If a node is not connected to any individual at the end of the algorithm, one link is added from that node towards a randomly chosen individual. 
The network structure is then described by the value of $\alpha_c$ and $\alpha_o$. We consider a network as random when $\alpha_c = \alpha_o$, and as centralized when $\alpha_c > \alpha_o$. \\

\subsection{Distribution of resources}
At each round, the group produces a fixed amount of resources $B = 2N$. The resources are distributed as a function of the social position $s_i$ of the individual, with the social position of the leader, clique and followers being respectively 0, 1 and 2. The fitness $w_i(t)$ of an individual $i$  at a time $t$ is equivalent to the resources received:
\begin{eqnarray}
w_i(t) = B*f(s_i(t))
\label{eqnFitness}
\end{eqnarray}
The function defining the distribution $f(s_i(t))$ is modulated by the leader preference $z_L$ such that :
\begin{eqnarray}
f(s_i(t)) =  \frac{e^{-s_i(t)*z_L(t)}}{\sum_{j=1}^{N}{-e^{s_j(t)*z_L(t)}}}
\label{eqnDistrib}
\end{eqnarray}
The distribution of resources is normalized and is bounded between an equal distribution of resources (with $z=0$) and a strongly asymmetrical distribution of resources (with $z=1$ ). We make the assumption that the leader has full control on collective decision. This is a common assumption in the literature on the evolution of despotism \citep{Buston2007ReproductiveModels}.

\subsection{Interactional justice}
Each individual $i$ has an opinion $o_i(t)$ describing its view of the current fairness of the society. It is the result of its own mindset $m_i(t)$, which is calculated from its own personal experience, and the mindset of its incoming social neighbours. 
First, an individual's mindset is calculated by comparing the resources it received with an egalitarian distribution:
\begin{eqnarray}
m_i(t) = \frac{1/N-p(t)}{1/N-p_{min}}
\label{eqnMindset}
\end{eqnarray}
The mindset is normalized by the difference between the maximum share $1/N$ and the minimum possible share $p_{min}$. As a result, the mindset is not dependant of the absolute amount of resources produced $B$. The opinion $o_i(t)$ of individual $i$ is then calculated as:
\begin{eqnarray}
o_i(t) = \frac{m_i(t)L + \sum_{j=1}^{k}{m_j(t)}}{k_i(t)+L},
\label{eqnOpinion}
\end{eqnarray}
with $j$ an incoming neighbour, $k_i(t)$ the in-degree of the focal node, and $L$ a weight determining the relative importance of its own experience compared to the mindset of neighbours. 
\\
The variables $m$ and $o$ are bounded between $0$ (totally satisfied) and $1$ (totally dissatisfied). An individual is considered defiant if its opinion value is more than its tolerance threshold $z_i(t)$. A defiant individual then pays a cost to attempt a revolution $C$.
In case of a large proportion of defiant individuals within the population, i.e. above a revolution threshold $T$, the current leader and clique become outgroup members and a new leader and clique are chosen from the defiant individuals. The network is then rebuilt.

\subsection{Reproduction}
We consider here the evolutionary process as only cultural evolution \citep{Boyd1985CultureProcess}. The evolution of the population is modelled by a Moran Process \citep{Moran1958RandomGenetics,Lieberman2005EvolutionaryGraphs}. This has been shown to be an efficient method to study evolution in finite populations and keeps the size of the population constant. The reproduction follows a death-birth process. At each time step, a randomly chosen individual dies. Then, the vacant node is replaced by the offspring of an individual chosen within the population with a probability proportional to its fitness, i.e. fitness-proportionate selection. The individual chosen to die is also competing to fill the vacant node with one of its own offspring. 
More formally, the new individual has a probability $P(i)$ to be the offspring of individual $i$ according to :
\begin{eqnarray}
P(i)(t) = \frac{w_i(t)}{\sum_{j=1}^{N}w_j(t)},
\label{eqnRepro}
\end{eqnarray}
with $N$ the population size, and $j=0$ the individual previously occupying the node. We assume that a vacant node can be replaced by any other individual in the population i.e. the individual changing its distribution preference can learn from the observation of any other individual.
Because we consider the opinion formation to happen on a longer time scale than the evolution of cultural items, this process is repeated $R$ times by generation.

\section{Results}
In this section we report our experimental results. To provide a comprehensive investigation of our research question, we perform two analyses. In our first analysis, we consider that only the leader expresses its distribution preference $z_L$ and that followers' distribution preference $z_f$ is fixed. Then, we combine mathematical analysis and numerical simulations to study the effect of the network structure e.g. $\alpha_c$ and $\alpha_o$ on the evolution of despotism. In our second analysis, we relax this assumption and use numerical simulations to allow both leader and follower preferences to evolve. We define the level of despotism as the level of inequality imposed by the leader which is here its distribution preference $z_L$.

\subsection{Analysis 1: Evolution of despotism level with fixed followers' tolerance}
We consider first that only the leader expresses $z$ and that followers' distribution preference $z_f$ is fixed. The fitness of the leader $w_L(t)$ is equal to:
\begin{eqnarray}
w_L(t) = \frac{1}{\sum_{i=1}^{N}{-e^{s_i*z_L(t)}}}.
\label{eqnFitLeader}
\end{eqnarray}
It can be shown that:
\begin{eqnarray}
\dv{w_L(t)}{z_L} = \frac{\sum_{i=1}^{N}{s_i e^{-s_i*z_L}}}{(\sum_{i=1}^{N}{e^{-s_i*z_L}})^2} > 0 
\label{eqnDFitLeader}
\end{eqnarray}
In other words, an increase in the level of despotism $z_L$ always increases the fitness of the leader and should be positively selected. However, it can exist a value $z^*$ of the leader trait $z_L$ between $0$ and $1$ for which the group undergoes a revolution. In this case, the leader becomes a follower and its trait $z$ no longer affects the distribution of resources. When the tolerance of followers is fixed, the distribution of resources is the only selection pressure existing on $z$. Consequently, the level of despotism $z_L$ will evolve towards the stable point $z^*$ defined as the maximum value of $z$ for which a revolution will not occur. 
The value of this evolutionary stable point is a function of the network structure, i.e. $\alpha_c$ and $\alpha_o$, for a given followers' tolerance and revolution threshold. Because it is not possible to analytically calculate $z^*$, we use numerical simulations to determinate its value as a function of $\alpha_c$ and $\alpha_o$.
 The default parameters used in the simulations, unless otherwise specified, are $N = 500$, $N_c = 25$, $L = 1$, $T = 0.1$, $z_f = 0.25$, $R = 100$. 
For each set of parameters considered, 100 independent simulations have been realized. The results presented, unless otherwise specified, are the mean value of replicates.

\begin{figure}[t]
\includegraphics[width=3.45in,keepaspectratio]{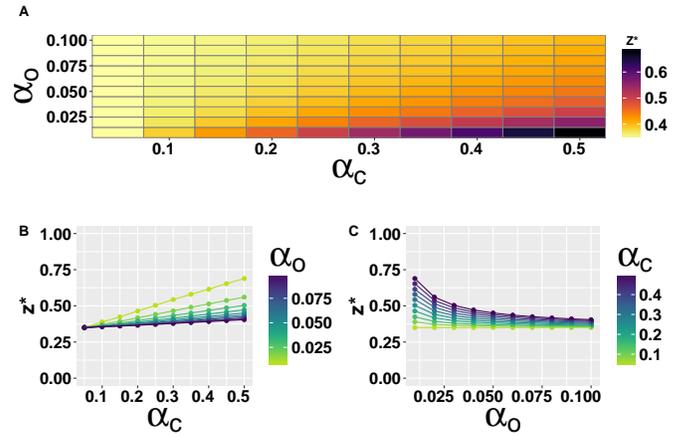}
\caption{Mean value of the evolutionary stable point $z^*$  in function of the connectedness of the clique $\alpha_c$ and connectedness of the outgroup $\alpha_o$}
\label{pt_zThrCO}
\end{figure}

Figure \ref{pt_zThrCO}.A demonstrates that centralization of the network structure leads to a higher level of despotism $z^*$. The greatest level of despotism $z^*=0.71$ is obtained for the maximum $\alpha_c$ and minimum $\alpha_o$; and the lowest level of despotism $z^*=0.35$ is obtained for the minimum $\alpha_c$ and maximum $\alpha_o$.
To better understand the contribution of each variable, a statistical model has been built. To take in account the heteroskedacity inherent to the model, we built a mixed linear regression model with $\alpha_o$ as a random effect. It shows that $\alpha_c$, $\alpha_o$ and their interaction have a significant effect on the level of despotism at equilibrium $z^*$ (p-value $< 1.10^{-6}$).
Because the presence of significant interaction limits the interpretation of the statistical model, a graphical representation is presented in figure \ref{pt_zThrCO}.B and figure \ref{pt_zThrCO}.C. They show that $\alpha_c$ has a linear positive effect on the level of despotism while $\alpha_o$ has a exponential linear effect on the level of despotism. In addition, it also depicts a strong interaction between the two variables with the positive effect of $\alpha_c$ on despotism being strongly dependent of the value of $\alpha_o$. In other words, a more centralized system lead to higher despotism only when followers are also disconnected from each other. Therefore, it suggests that increasing connectedness of outgroup members is a efficient way to limit the evolution of despotism.

\subsection{Analysis 2: Evolution of despotism level and follower's tolerance}
In our second analysis, we allow the tolerance of followers to evolve. Because of the complexity of the model, we use numerical simulations to analyse the model. The results of interest are the mean value of the distribution preference $z$ and the level of despotism defined as the leader's value of distribution preference $z_L$. In addition, we present the mean value of mindset $m$ obtained from self-assessment, the mean value of opinion $o$ obtained from interaction with neighbours, the mean value of bias defined as the difference between $m$ and $o$ and the frequency of revolution events within the population. Because of the presence of a mutation operator, the system is an ergodic Markov Chain with no absorbing states. Therefore, the result of interest is the stationary distribution, which describes how long the system spends in each state. To calculate this, we present the average over long-run time over $5X10^7$ generations by sampling $50$ data points every $1X10^6$ time steps. This method is confirmed as a good approximation of the stationary distribution by the absence of a periodic pattern of cycles and the standard error between simulations being always less than $0.027$ .
The default parameters used in the simulations, unless otherwise specified, are $N = 500$, $N_c= 25$, $L = 1$, $T = 0.1$,  $R= 100$,  $C = 0.1$, $\sigma_\mathrm{m} = 0.01$ and $\mu = 0.01$. The initial values of $z$ are randomly generated. For each set of parameters considered, $50$ independent simulations have been realized. The box plots represents the dispersion of the mean value across time. The results presented as scatter plots show the mean value of replicates and the error bars represent the standard error between the mean value of replicates.

\begin{figure}[t]
\includegraphics[width=3.45in,keepaspectratio]{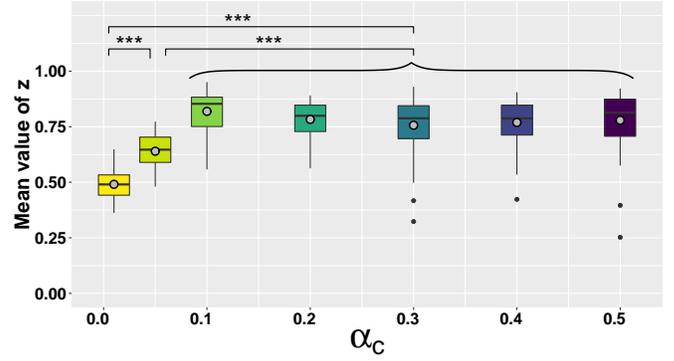}
\caption{Long-run time averages over $5$ X $10^7$ generations of the mean level of fairness $z$ as a function of clique connectedness $\alpha_c$. Grey circles represent the mean value of distribution preference of the leader $z_L$. Results are compared by pairwise Welch's t-test (***: p-value $< 1X10^{-6}$)
}
\label{pt_ZAlphaC}
\end{figure}

\begin{figure}[t]
\includegraphics[width=3.45in,keepaspectratio]{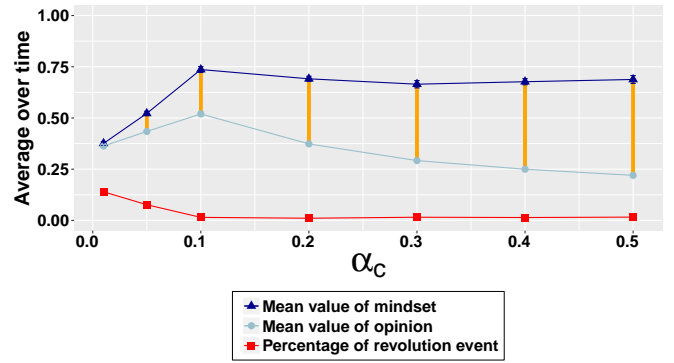}
\caption{Long-run time averages over $5$ X $10^7$ generations of the mean mindset, mean opinion $z$ and mean percentage of revolution as a function of clique connectedness $\alpha_c$. Orange bars represent the mean value of bias defined as the difference between mindset and opinion.
}
\label{pt_MOR_AlphaC}
\end{figure}

Figure \ref{pt_ZAlphaC} confirms that increasing the connectedness of the leader and its clique significantly leads to a higher level of despotism $z_L$, even when the distribution preference of followers also evolves. Figure \ref{pt_ZAlphaC} shows that above $0.1$, further increasing the connectedness of the clique does not have a significant effect. However, this plateau is explained by the maximum limit imposed on the distribution preference $z$. As before, this result is explained by the evolution of leader distribution preference being controlled by the threshold at which followers start a revolution and change the leader and its clique. Figure~\ref{pt_MOR_AlphaC} highlights the mechanism behind the centralization effect: an increase in clique connectedness $\alpha_c$ is translated into a higher negative bias of opinions which leads to a lower frequency of revolution, and ultimately a higher mean level of despotism.
In addition, Figure \ref{pt_ZAlphaC} demonstrates a similar positive effect of the connectedness of the leader and its clique on the mean value of distribution preference $z$. In other words, centralization lead followers to be more tolerant to despotism. By deciding of a more skewed distribution of resources, the leader increases its fitness which causes its distribution preference to spread in the population. This effect associated with the cost of revolution leads to the mean value of distribution preference being close to the leader distribution preference.
It is also worth noting that even in a random network and in absence of bias, followers evolve a relative tolerance to despotism. In addition, in contrast to the previous result, the model including the evolution of followers' preference has an overall higher level of despotism. This result is explained by the follower preference for equality being limited by the cost of revolution and the necessity of having a threshold proportion of individuals being in a defiant state at the same time. 
Finally, a close-up look at the simulations show that $z$ strongly vary because of succession of period of increasing despotism and period of revolution. Indeed, the follower preference for equality is dependent of the leader preference and leads to chaotic variations. Despite this, the upper limit value of $z$ and its average on long-run time confirms the positive effect of centralization on the level of despotism.

\begin{figure}[t]
\includegraphics[width=3.45in,keepaspectratio]{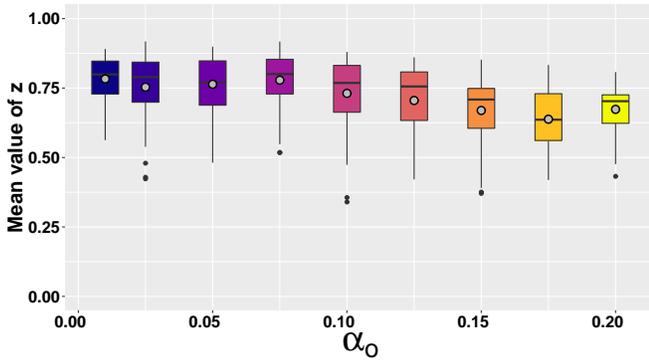}
\caption{Long-run time averages over $5$ X $10^7$ generations of the mean level of fairness $z$ as a function of outgroup connectedness $\alpha_o$. Grey circles represents the mean value of distribution preference of the leader $z_L$. 
}
\label{pt_ZAlphaO}
\end{figure}

\begin{figure}[t]
\includegraphics[width=3.45in,keepaspectratio]{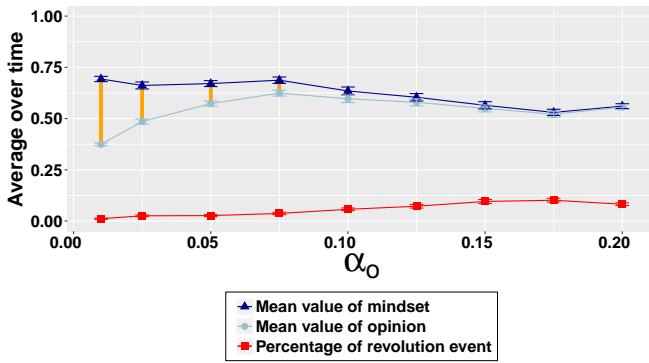}
\caption{Long-run time averages over $5$ X $10^7$ generations of the mean mindset, mean opinion $z$ and mean percentage of revolution as a function of outgroup connectedness $\alpha_o$. Orange bars represent the mean value of bias defined as the difference between mindset and opinion.
}
\label{pt_MOR_AlphaO}
\end{figure}

Figure \ref{pt_ZAlphaO} shows that increasing the connectedness of followers leads to a lower level of despotism $z_L$, even when the distribution preference of followers also evolves. In addition, it demonstrates a similar effect on the mean value of distribution preference $z$ for the reason stated previously. Both of these effects have been tested using a linear regression and are statistically significant (p-value $< 1.10^{-6}$). 
It is worth noting that the effect of the connectedness of followers on the level of despotism is smaller in comparison to the results where the tolerance of followers is fixed. However, Figure~\ref{pt_MOR_AlphaO} confirms that increasing the connectedness of followers greatly reduces the bias and therefore increases the frequency of revolution in response to despotic behaviour. In other words, the influence of the leader and its clique which blind followers judgement is dependent of disconnected followers. Therefore, the smaller effect of $\alpha_o$ on the level of despotism in this analysis is due to the other constraints affecting the cost and benefit of revolution as stated in the results looking at the effect of $\alpha_c$ . It suggest that the revolution mechanism also affects the evolution of despotism and should be investigated in future work.

\section{Discussion}

\subsection{Summary}
Despite the potential benefits of the hierarchy, centralization appears to go along with despotism, i.e. inequality enforced by leaders. Yet, it is still hard to determine if inequality and despotism are an inherent consequence of centralization or the result of a common element. Although different factors have been identified, the role of distribution of influence and its impact on knowledge transfer has not yet been investigated. To fill this gap, we have simulated such a scenario by modelling the evolution of distribution preference in artificial societies structured in different social networks. 
The model developed demonstrates that the centralization of social networks leads to the evolution of higher despotism and inequality. In other words, an asymmetrical distribution of influence is sufficient to create inequality. This result holds when the tolerance of followers is fixed or in a more realistic set-up where tolerance of followers evolve. This result is explained by the knowledge of followers on the leader's decision which is (i) biased by the influential members of the clique, and (ii) limited by their low connectedness to other followers. As a consequence, followers can't impartially enforce their control on leaders and \textit{a fortiori} on collectively decided institutional rules. 
Furthermore, the model demonstrates that the effect of influential members on followers' opinion is strongly dependent on followers having low connectedness. Indeed, a slight increase in the influence of followers greatly reduces the despotism created by the clique influence. However, as shown by comparing the first and second analysis, this effect is weaker when follower's distribution preference is also evolving. Overall, this result suggests that increasing the connection between followers could be a solution to limit despotism in social systems.

\subsection{Related work}

The results presented here attempt to bridge the gap between two main research axes. Previous research work has either examined the impact of centralization on opinion formation processes, but without evolutionary processes, or has studied the evolution of despotism without integrating mechanisms underlying opinion formation. 
On the former side, \citet{Gavrilets2016ConvergenceLeadership} have shown  that the presence of highly influential individuals can strongly bias the collective decision. Later on, \citet{Pitt2017InteractionalSystems} has integrated institutional rules and interactional justice into a multi-agent systems and shown that hierarchy can bias the followers' opinion on leader decided rules. We have here confirmed that this result still holds even when the evolution of individual preferences for the distribution of resources are taken in account. Furthermore, we have shown that integrating the evolution of followers' preferences can lead to irregular level of despotism but yet, with the same qualitative behaviours as when only the leader's distribution preference evolves.
On the other hand, reproductive skew theory used mathematical models to understand how the conflict between leader and follower affects the evolution of despotism \citep{Summers2005TheDespotism}. These models have identified the important factors behind the evolution of inequality such as the cost of leaving the group or the relative cost of conflict with the leader. Later on, evolutionary models have extended and confirmed this work by taking in account many connected large groups \citep{Hooper2010AGroups,Powers2014AnDespotism}. Our results complete this previous work by integrating an opinion formation process and by identifying a new crucial factor in the evolution of inequality: the distribution of influence itself as modelled by social network structure. 

On the whole, providing a complete explanation of the evolution of group organization and complexity is a crucial goal in research \citep{Szathmary2015Toward2.0}. The ``iron law of oligarchy'' provides such a scenario by stating that (i) hierarchy emerges when groups grow bigger as a consequence of egalitarian organization becoming more costly, and (ii) despotism appears as a consequence of hierarchy distributing political power asymmetrically. However, this scenario is missing a micro-level explanation \citep{Mitchell1915PoliticalDemocracy}. The model developed here has shown that an asymmetric distribution of influence is sufficient to lead to despotism and inequality. Along the same line, previous work has shown that this asymmetric distribution of influence tends to evolve in large groups to reduce the cost of organization \citep{Perret2017EmergenceModel}. Taken together, this work draws a first outline of a mechanistic description of the iron law of oligarchy. 

Finally, our model predicts that the capacity of followers to efficiently control leader's decision is crucial to limit despotism. This result is supported by evidence from behavioural economics experiments. In particular, two economics games called the ultimatum and the dictator game implement a similar version of the presented model. In the ultimatum game, one of two players has to decide how to split a fixed amount of money and the second player can choose to either accept it and both receive their shares; or refuse it in which case neither receive anything. However, in the variant called dictator game, the second player can't decide to accept or refuse. Experimental results show that in the ultimatum game, the proposer keeps in average 60\% of the total amount while in the dictator game the share kept by the proposer goes up to 72\% \citep{Oosterbeek2004CulturalMeta-analysis,Engel2010DictatorStudy}. In conclusion, the ability of followers to punish the leader reduces its tendency toward despotism. Importantly, our results show that centralization of the social network can blind the judgement of followers and transform the distribution of resources from an ultimatum game to a dictator game.

\subsection{Assumptions and further work}
We made a number of assumptions to keep our model tractable. First of all, the model developed considers only blind evolutionary processes as a driver of change in distribution preferences. However, cognitive processes might also affect the evolution of agent preferences and lead to a lower level of despotism e.g. followers predict that a low level of despotism favour their positions. This difference suggests that integrating cognitive processes might be crucial to limit despotism in artificial social systems and would be worth investigating. Yet, it is important to note that our results still hold over large time scales in which blind evolutionary processes are a good predictor of cultural change \citep{Boyd1985CultureProcess}. Another assumption made concerns the division of the society into only three groups and with only one leader. In natural social systems, hierarchy can be composed of many more layers. But this is unlikely to change our qualitative results since the results presented are explained by the asymmetrical distribution of influence. Nonetheless, it is crucial to explore similarly the evolution of despotism in other hierarchical network structures. Finally, we have considered here a simplified version of the revolution process. \citet{Weingast1997TheLaw} have used a game theory model to show that the cost of coordination required to make a revolution would lead to either a fair society or a strongly despotic one. In our model, this limit induced by coordination is represented by the minimum amount of defiant individuals required to start a revolution. Similarly we find that society can switch from equality to strong despotism, but we also show that groups might vary widely along this range. In addition, our results demonstrate that this effect is affected by the structure of the social network. However, extending the current model to integrate more explicitly revolution as a Volunteer's Dilemma game, along with individual strategies for playing this game, could provide a better insight on the evolutionary dynamics of despotism. In particular, it could clarify the impact of follower's connectedness on the evolution of despotism.

\subsection{Conclusion}

Social systems organised in hierarchy tend to develop into despotism with inequality created by and for the leaders. It has been proposed that the sole asymmetrical distribution of power is enough to lead to such transition. In particular, the leader and its influential clique could bias the opinion of weakly connected followers, ultimately crippling the followers' capacity to control the leader's decision. Yet, this scenario was missing of a model integrating and testing explicitly these mechanisms. To fill this gap, we have developed a model integrating evolutionary processes, opinion formation and interactional justice. We have used numerical simulations to investigate the impact of centralization of the social network on the evolution of the level of despotism. Our results have demonstrated that the centralization of a social network would lead to higher despotism and inequality. It predicts that a transition from equality to despotism will happen in presence of (i) highly influential individuals with a preferential access of resources; and (ii) lowly connected followers. In addition, our model demonstrates how a low-level process such as opinion formation can strongly drive the evolution of a higher property, here the group organization.

\footnotesize
\bibliographystyle{apalike}
\bibliography{Mendeley.bib}

\end{document}